# Intrinsic Point Defects in Ultrathin Layered 1T-PtSe$_2$

Husong Zheng[1†], Yichul Choi[1†], Fazel Baniasadi[1,2†], Dake Hu[3†], Liying Jiao[3*], Kyungwha Park[1*], Chenggang Tao[1*]

[1]Department of Physics, Virginia Tech, Blacksburg, Virginia 24061, USA

[2]Department of Materials Science and Engineering, Virginia Tech, Blacksburg, Virginia 24061, USA

[3]Department of Chemistry, Tsinghua University, Beijing 100084, China

[†]These authors contributed equally to this work.

[*]Corresponding Authors: lyjiao@mail.tsinghua.edu, kyungwha@vt.edu and cgtao@vt.edu.

**ABSTRACT**

Among two-dimensional (2D) transition metal dichalcogenides (TMDs), platinum diselenide (PtSe$_2$) stands at a unique place in the sense that it undergoes a phase transition from type-II Dirac semimetal to indirect-gap semiconductor as thickness decreases. Defects in 2D TMDs are ubiquitous and play crucial roles in understanding and tuning electronic, optical, and magnetic properties. Here we investigate intrinsic point defects in ultrathin 1T-PtSe$_2$ layers grown on mica through the chemical vapor transport (CVT) method, using scanning tunneling microscopy and spectroscopy (STM/STS) and first-principles calculations. We observed five types of distinct defects from STM topography images and obtained the local density of states of the defects. By combining the STM results with the first-principles calculations, we identified the types and characteristics of these defects, which are Pt vacancies at the topmost and next monolayers, Se vacancies in the topmost monolayer, and Se antisites at Pt sites within the topmost monolayer. Our study shows that the Se antisite defects are the most abundant with the lowest formation energy in a Se-rich growth condition, in contrast to cases of 2D molybdenum disulfide (MoS$_2$) family. Our findings would provide critical insight into tuning of carrier mobility, charge carrier relaxation, and electron-hole recombination rates by defect engineering or varying growth condition in few-layer 1T-PtSe$_2$ and other related 2D materials.

**Subject Areas:** Condensed Matter Physics

# I. INTRODUCTION

Transition-metal dichalcogenides (TMDs) with a general formula of $MX_2$, where M represents transition metal elements (groups 4-10) and X represents chalcogen elements (S, Se or Te), are a family of two-dimensional (2D) materials being extensively studied in the past few years [1-4]. A single TMD layer consists of a hexagonal layer of the M atoms sandwiched between two hexagonal layers of the X atoms. Neighboring TMD layers are typically coupled via a weak van der Waals interaction. Depending on the number of *d* electrons and thickness, TMDs can have a variety of electronic properties, namely metallic, semimetallic, semiconducting and superconducting [5,6]. Physical and chemical properties of these 2D materials are markedly different from their bulk counterparts and can be tuned for wide ranges of applications [7-10].

So far, studies on TMDs have been mostly conducted on $MX_2$ with group VIB transition metals, such as M = Mo, W, and X = S, Se. Recently a new type of TMD, platinum diselenide ($PtSe_2$) in a 1T structure (FIG. 1a), has been synthesized in bulk form and ultrathin layers [2,11-13]. Compared to the well-studied TMDs, 1T-$PtSe_2$ has inversion symmetry and it has stronger coupling between neighboring unit layers. Furthermore, this material is unique in the sense that a transition from indirect-gap semiconductor to metal can be driven by simply varying thickness [14]. Among all the TMDs, $PtSe_2$ has the highest Seeback coefficient good for thermoelectric applications [15] and the extremely high mobility, up to 3000 $cm^2/V/s$, desirable for electronic applications [16]. $PtSe_2$ can be also used in catalysis [2,17] and as efficient gas sensors because of low adsorption energies for gases like NO, CO, $CO_2$ and $H_2O$ [5]. In monolayer 1T-$PtSe_2$ layer, spin polarization induced by a local Rashba effect was recently observed [13].

Defects are ubiquitous in 2D TMDs, especially those synthesized via chemical vapor deposition (CVD) or transport (CVT) [18,19]. Some defects appear from growth or annealing processes [20], whereas some other defects are naturally or intentionally brought into the structure during investigation [21]. In 2D materials, typical zero-dimensional or point defects constitute vacancies, antisites, adatoms, intercalations, interstitial dopants, and substitutional dopants [4,22], while one-dimensional defects include grain boundaries, dislocations, and edges [23-25]. Properties of 2D TMDs are very sensitive to defects, especially for 2D semimetals and semiconductors [21,26-29]. Depending on the properties of interest and desirable applications, defects can be beneficial or detrimental. For instance, 60° twin grain boundaries in molybdenum and tungsten dichalcogenides can function as metal wires (conducting pathways) or sinks for carriers [30]. Point defects typically lower the carrier mobility or degrade mechanical properties of 2D materials [31]. On the other hand, under certain conditions, point defects can also be sources of single photon emission [3,32] and induce large spin-orbit splitting in 1T TMDs [33,34].

Despite the ubiquity and importance of defects, there are no experimental studies of defects in 1T-$PtSe_2$ layers at the atomistic level yet. Here we investigate intrinsic point defects for ultrathin 1T-$PtSe_2$ layers grown on mica through the CVT method, by using scanning tunneling microscopy and spectroscopy (STM/STS) and first-principles calculations. Point defects were formed in the structure of 1T-$PtSe_2$ during the growth process. As shown in FIG. 1a, in a 1T structure, $PtSe_2$ monolayers are stacked in a fashion of A-A. Through STM/STS, we identified five types of dominant point defects and obtained their atomic structures and local density of states. We determined characteristics and formation energies of the defects by using density-functional theory (DFT). Our results may stimulate studies of effects of defects on electronic and optical properties and defect engineering for applications in this interesting new ultrathin 1T-PtSe$_2$ and other 2D TMDs.

## II. METHODS

### A. Experiment

Ultrathin PtSe$_2$ flakes were grown on a mica substrate by the CVT method at a growth temperature of 600-700 °C. The detailed procedures were reported in our previous papers [35,36]. Then a stripe of 100-nm thick gold film was evaporated through a shadow mask on the samples as electrodes. The samples were annealed at 250 °C for 2.5 hours in the preparation chamber in a customized Omicron LT STM/AFM system with a base pressure of low $10^{-10}$ mbar before transferring it into the STM analysis chamber that is connected to the preparation chamber. In the main text, all the STM and STS results were carried out in the customized Omicron STM/AFM system and all the measurements were performed at 77 K. Room-temperature STM data are provided in the Supplementary Information. STM imaging was carried out at a constant current mode, and STS measurements were done at an open feedback loop using a bias modulation 20 mV with the frequency of 1000 Hz.

### B. Simulation

DFT-based simulations were performed including spin-orbit coupling (unless specified otherwise) using VASP [37,38]. We employed local-density approximation (LDA) [39] for the exchange-correlation functional and used projector-augmented wave (PAW) pseudopotentials [40]. We chose LDA because it gives both in-plane and out-of-plane lattice constants of bulk PtSe$_2$ closer to the experimental values [41-43] than the Perdew-Burke-Ernzerhof (PBE) generalized-gradient approximation (GGA) [44] upon geometry relaxation, as reported in Refs. [45,46]. The PBE-optimized out-of-plane lattice constant $c$ for bulk is 27-29% larger than the experimental value, as shown in the literature [45,46]. Furthermore, with the PBE-optimized $c$ value, a band gap opens for bulk $PtSe_2$ [45], which is inconsistent with the experimental observation [47,48]. There are also conflicting results [45,46] in the improvement of the out-of-plane lattice constant when van der Waals interaction [49,50] is included within PBE-GGA.

A surface of a thin $PtSe_2$ film was modeled by two 1T-$PtSe_2$ monolayers. Supercells of 5×5 surface atoms were used except for the Pt2 vacancy defect (discussed later), in order to simulate isolated defects. The energy cutoff was set to 260 eV and a k-point mesh of 5×5×1 was used. We fully relaxed the supercell structures until the residual forces were less than 0.01 eV/Å. For the Pt2 vacancy defect, a 7×7 in-plane supercell was used to simulate large-area modulations of the observed STM image near the defect site. Only in this case, spin-orbit coupling was turned off, and an energy cutoff of 230 eV and a k-point mesh of 3×3×1 were employed. For the slab calculations, we included a vacuum layer thicker than 20 Å to avoid interactions between successive images of $PtSe_2$ layers.

To compare with the STM topographic images, applying Tersoff-Hamman approach [51,52], we integrated the DFT-calculated surface local density of states (LDOS) from the Fermi level to the experimental bias voltage at a plane $z = 1$ Å above the topmost atomic layer. Although isosurfaces of the integrated LDOS corresponding to constant current are more accurate, the method we applied has been used as a good approximation to STM topographic images in the literature [53-55]. The integrated LDOS images were visualized using VESTA [56]. We also calculated formation energies of various defects using the standard method discussed in Ref. [57]. A brief description of the method is shown in the Supplementary Information.

# III. RESULTS AND DISCUSSION

## A. STM topographic images

The atomic structure of 1T-PtSe$_2$ is schematically drawn in FIG. 1a. We used atomic force microscope (AFM) and STM to determine the number of layers of PtSe$_2$ flakes (FIGs. 1b, S1 and S2). Considering the reported thickness of a single PtSe$_2$ layer, 0.507 nm [42], we determined that the thickness of flakes in our measurements ranges from 5 to 9 layers. A high-resolution STM topographic image of a defect-free 1T-PtSe$_2$ surface is shown in FIG. 1c. Similar to previous STM study on monolayer PtSe$_2$ grown though selenization of a Pt(111) substrate and other transition-metal diselenides, the hexagonal protrusions in FIG. 1c represent the topmost Se atoms [2,58]. From the arrangement of the surface Se atoms, the in-plane lattice constant of 1T-PtSe$_2$ is determined to be $0.375 \pm 0.003$ nm, which is consistent with previous theoretical and experimental results [2,17,59]. FIG. 2 shows a typical atomically resolved large scale STM image showing various defects appearing on the PtSe$_2$ surface. FIGs. 2a and 2b are the STM images of the same area obtained at positive (empty states) and negative sample bias voltages (filled states), respectively. These defects can be visualized at the surface with distinct atomic structures and morphologies at a given bias voltage, marked by arrows in FIG. 2. The morphology of some types of defects reveals a clear dependence on bias voltage. Combining the defect morphology at the filled and empty states, we identified five dominant types of point defects at the PtSe$_2$ surface, labeled as A, B, C, D, and E, as shown in FIG. 2. In a single-crystalline flake, we noticed that each defect type has the same orientation. In addition to these five defect types, we occasionally observed a few other defects which can be either combined or new defects. For simplicity, we henceforth focus on these five dominant types of defects.

FIGs. 3a-e show zoom-in STM images of the five defect types. Defects of type A appear like depletions at both positive and negative bias voltages. Defects of type B, C, and D look like protrusions at both positive and negative bias voltages. For defects of type E, protrusions (depletions) are shown at negative (positive) bias voltages. All these types of defects have three-fold symmetry. Defects of type A are centered at $Se_{top}$ sites, while defects of type B, C, D, and E are centered at Pt, $Se_{bottom}$, Pt and Pt sites, respectively, where $Se_{bottom}$ ($Se_{top}$) is the bottom (top) Se site of the topmost monolayer, as marked by the dashed triangles in FIG. 3 (see also FIG. S3 in the Supplementary Information). Table I lists the density of these defects obtained by averaging over STM images of typical regions each with an area of $20 \times 20$ nm$^2$. Among these five types, defects of type E have the highest density. We now discuss each defect type separately.

Defects of type A show one-site depletions at the surface at both positive and negative bias (FIG. 3a). There is not a much difference in the brightness of the STM images at various bias voltages. This defect type could be due to a missing Se atom on the surface of the topmost $PtSe_2$ layer, in other words, a Se vacancy in the top Se layer, labeled as $V_{Se1a}$.

Defects of type B show $1 \times 1$ triangular protrusions at both positive and negative bias voltages, although the protrusions are much more apparent at positive bias voltages (FIG. 3b). There are a few factors affecting tunneling current, such as the height of the atoms at the surface and integrated local density of states [52,60]. Typically, the higher surface atoms or larger integrated local density of states result in protrusions in STM images. Despite this difficulty, our observation suggests that the defects of type B are electron acceptor defects. Since we cannot determine the orientation of 1-T structure form the top Se layer, each of the three sites may be associated with either (i) Se substituted by a more electronegative element at the bottom atomic layer of the topmost monolayer, or (ii) a Pt vacancy ($V_{Pt1}$) located at the topmost monolayer, noted as Pt1 in FIG. 1a.

Option (i) is, however, ruled out since there were no other observable more electronegative elements than Se in the growth environment. Option (ii) is more promising since the growth was under a Se-rich condition. The Pt vacancy can build a high acceptor density at the three neighboring Se sites forming the triangle, giving rise to the three protrusions at positive bias. This analysis is supported by our DFT calculations, as explained later.

Defects of type C show 2 × 2 triangular protrusions at both positive and negative bias voltages (FIG. 3c), although the protrusions are much more apparent at positive bias voltages, similarly to defects of type B. Considering that the defects of type C are centered at the bottom Se atoms, we come up with three candidates for the origin of such a defect, such as a bottom Se atom replaced by Pt ($Pt_{antisite}$, $Pt_{Se1b}$), a bottom Se atom vacancy ($V_{Se1b}$), or three neighboring Pt vacancies around the center bottom Se atom ($V_{3Pt}$).

Defects of type D show 4 × 4 triangular protrusions at both positive and negative bias voltages (FIG. 3d), and they are centered at Pt sites just like the defects of type B. Protrusions are more prominent at the vertices of the triangles, but they are not as strong as those for the defects of type B and C at positive bias voltages. Compared with the defects of type B and C, topographic images do not show much contrast between positive and negative bias voltages.

Defects of type E appear as 1 × 1 triangular protrusions at negative bias, and show the same triangular protrusion surrounded by depletions at positive bias (FIG. 3e). Among the five defect types, this type appears with the highest density (Table I), which is consistent with the DFT-calculated formation energies of defects discussed later. This finding in 1T-$PtSe_2$ layers is in contrast to the trend of defects in group VIB TMD $MoS_2$ family layers where vacancies are the most dominant defects in either Mo- or S-rich condition [3].

## B. DFT simulations of integrated LDOS

In order to determine the characteristics of the five defect types, we performed DFT calculations. We considered all possible intrinsic single point defects within a PtSe$_2$ slab of two monolayers as well as Se adsorption based on the Se-rich growth condition. Within each atomic layer, two types of single point defects, vacancy and antisite, were considered. In addition, intercalation of a Se or a Pt atom within the van der Waals gap was taken into account. Although the experimental sample flakes are about five to nine monolayers thick, such thick slabs including defects cannot be simulated due to high computational cost. We calculated the total DOS for the pristine bilayer and six monolayer using the LDA-optimized geometries. The overall features of the total DOS for the bilayer are similar to those for the six monolayer, as shown in Fig. S5 in the Supplementary Information.

FIG. 4 shows integrated LDOS images for five distinct types of defects which are closest to the experimental STM images of defect types A-E at positive and negative bias voltages. These images as well as our calculated formation energies (Sec. III. C) bolster the identification of each observed defect type. According to the DFT-calculated LDOS images, defect types A-E correspond to $V_{Se1a}$, $V_{Pt1}$, $V_{Se1b}$, $V_{Pt2}$ and $Se_{Pt1}$, respectively.

Defects of type A are indeed from one missing atom at the topmost Se layer, $V_{Se1a}$, clearly seen in both experimental images and integrated LDOS (FIGs. 3a and 4a). In various TMDs [4,58,61], defects of this type are the most abundant and produce electron trap states within the band gap.

For defects of type B, our integrated LDOS images show that $V_{Pt1}$ produces 1 × 1 triangular protrusions at positive and negative bias (FIG. 4b), similarly to the experimental STM images. We also found that after the structural relaxation, three nearest neighboring Se atoms at the topmost

Se layer around the Pt vacancy site are expanded outward in plane (~ 0.3 Å) and vertically (~ 0.1 Å) due to the missing bonds between the Pt and Se atoms. This topographic effect explains why the STM images for defects of type B show 1 × 1 bright triangular shapes at both negative and positive bias as well as the small expansion of the in-plane lattice constant in the triangular protrusions.

For defects of type C, among the three aforementioned possibilities ($Pt_{Se1b}$, $V_{Se1b}$, $V_{3Pt}$), we found that $V_{Se1b}$ produces the 2 × 2 triangular protrusions similar to those observed in the STM images at positive and negative bias (FIG. 4c). Unlike the case of type B defects, changes in the atomic positions near the defect site were negligible upon the structural relaxation. Therefore, the observed STM image is likely to be determined by the influence of the defects on the nearby electronic structure, rather than by mere topographic changes.

The integrated LDOS images confirm that defects type D arise from $V_{Pt2}$. From the simulations, we observe 4 × 4 triangular protrusions at positive and negative bias voltages (FIG. 4d). The three Se atoms at the vertices of the triangle are vertically shifted upward by about 0.04 Å, which explains moderately bright protrusions at both positive and negative bias.

The observed STM images for defect type E agree with the integrated LDOS images of $Se_{Pt1}$ defects (FIG. 4e), although the effect of defect type E is quite subtle. Qualitative features such as 1 × 1 triangular protrusions at both negative and positive bias voltages and the surrounding depletions at positive bias are close to those of the experimental images.

### C. DFT calculations of defect formation energies

For a better understanding of stability of different defect types, we calculated the formation energies of all the defect types we considered in this work, as shown in FIG. 5. The formation energy values at the extreme Se-rich and Pt-rich conditions for all defect types are listed in Table S1 in the Supplementary Information. Since we calculated the formation energies for a $PtSe_2$ bilayer with inversion symmetry, Pt1 and Pt2 atoms are equivalent to each other and the formation energies of $V_{Pt1}$ and $V_{Pt2}$ are the same in our calculations. The formation energies of some defect types are not shown in FIG. 5 but listed in Table S1, because their formation energies are much higher than the other observed defects. Our calculated formation energies agree with the STM experimental densities of defects listed in Table I, although the experimental sample flakes are thicker than a bilayer. For instance, in a Se-rich condition, the formation energy of $Se_{Pt1}$ defect alone is well below the formation energies of other defects, becoming negative in the extreme Se rich case. In our sample grown under a Se-rich condition, defect type E turns out to have the highest density as observed in experiments (Table I), and that the integrated LDOS images (FIG. 4e) suggest that defect type E arises from the $Se_{Pt}$ antisite defect. In addition, Table I shows that the density of defect type A is higher than that of defect type C. This agrees with our result that the formation energy of defect type A is noticeably lower than that of defect type C (FIG. 5).

Some defect types shown in FIG. 5 and Table S1 have low formation energies but they have not been observed in the STM experiment. For example, Se adatoms (at the hollow Se site, hollow Pt site, and on top site) have not been observed, although their formation energies are significantly lower than those of the observed defect types. This could be due to the annealing process done before transferring into the STM analysis chamber.

### D. STS analysis

To characterize the electronic structures near the defects, we performed STS measurements on both defect-free and defect-rich areas. FIG. 6 shows dI/dV spectra of pristine areas and defect-rich areas with types A-E with a thickness of six monolayers, in the voltage range of $-0.8$ V to 0.8 V.

For the defect-free areas the V-like shaped dI/dV spectrum has much higher density of states (DOS) at empty states than at filled states with a rather wide region of low DOS near the Fermi level. This feature indicates that pristine PtSe$_2$ is semi-metallic at the thickness down to six layers, which is consistent with the reported experimental data [47]. The dI/dV spectra of the defects were measured at the center for defect type A and at the protrusions of defect types B, C, D and E, respectively. We also measured the STS at the centers of the defects, and found less prominent features compared with those measured at the protrusions, especially for V$_{Pt2}$ defects. Each dI/dV curve was obtained by averaging over 15 spectra. Due to the metallic nature of the PtSe$_2$ flakes, the dI/dV peaks for the defect-rich regions are not as prominent as those for typical semiconducting layers with defects.

## IV. CONCLUSION

We investigated intrinsic point defects in ultrathin PtSe$_2$ layers grown via the CVT method, by using STM/STS and first-principles calculations. We observed and identified five dominant types of point defects, such as V$_{Se1a}$, V$_{Pt1}$, V$_{Se1b}$, V$_{Pt2}$ and Se$_{Pt1}$. We calculated the formation energies of these defect types and compared them with the densities of the defects observed in experiments. The relative densities of the dominant defect types are in good agreement with the calculated formation energies. The experimental data and theoretical results suggest that Se$_{Pt1}$ antisite defects are the most abundant with the lowest formation energy in the Se-rich condition. Our findings elucidate the modification of electronic structures from the point defects, which would be crucial for optimizing the growth of ultrathin PtSe$_2$ layers and designing future electronic and spintronic devices.

# AUTHOR INFORMATION

**Corresponding Authors**

*Email: (L.J.) lyjiao@mail.tsinghua.edu, (K.P.) kyungwha@vt.edu, and (C.T.) cgtao@vt.edu.

**Author Contributions**

C.T. conceived and designed the research. H.Z., F.B. and C.T. performed STM measurements and analyzed STM data. D.H. and L.J. synthesized $PtSe_2$. Y.C. and K.P. carried out theoretical simulations. All authors discussed the results and wrote the paper.

## ACKNOWLEDGMENTS

C.T. and H.Z. acknowledge the financial support provided for this work by the US Army Research Office under the grant W911NF-15-1-0414. L.J. and D. H. acknowledge National Natural Science Foundation of China (No.51372134, No.21573125) and Beijing Municipal Science & Technology Commission (No.Z161100002116030). Y.C. was supported by the Virginia Tech Institute for Critical Technology and Applied Science (ICTAS) fellowship. The computational support was provided by San Diego Supercomputer Center (SDSC) under DMR060009N and Virginia Tech Advanced Research Computing (ARC).

**FIGURES**

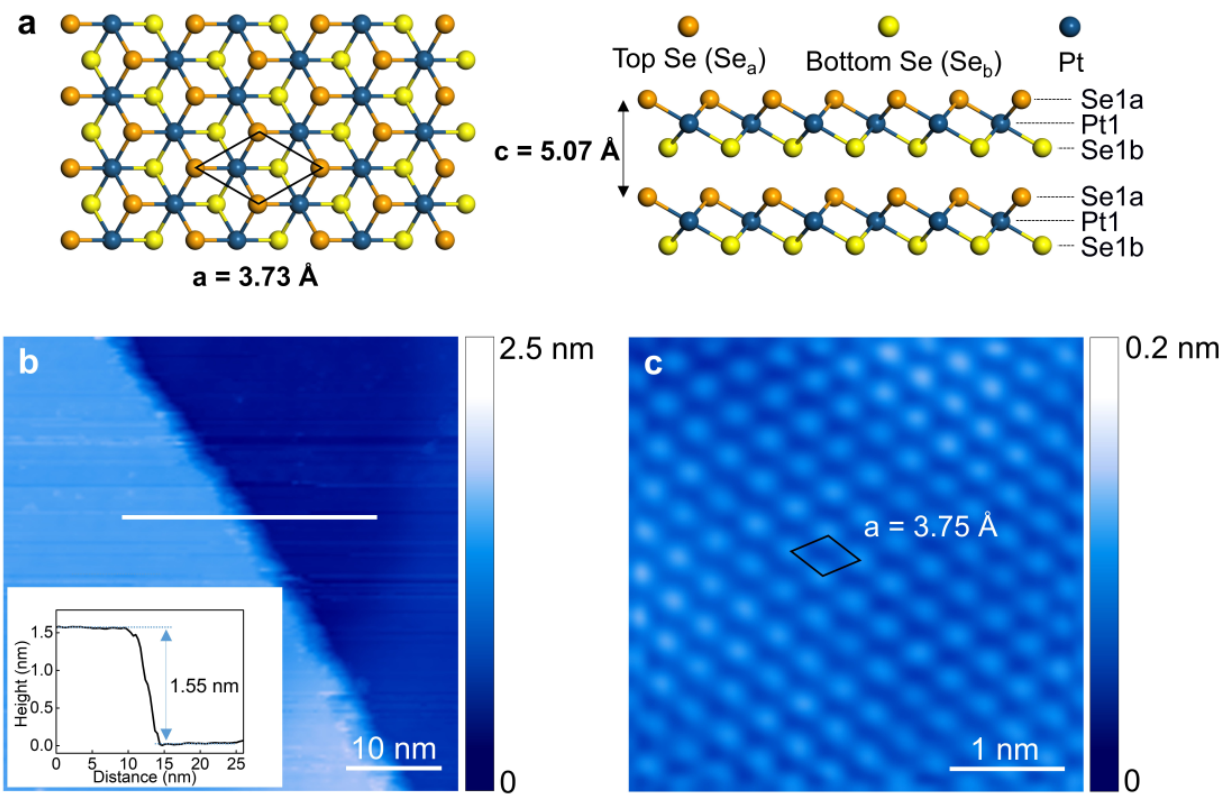

**FIG. 1.** (a) Structure model of 1T-PtSe$_2$. (b) Large scale STM image of a few-layer PtSe$_2$ flake, including a step edge indicating that the left area is three-layer thicker than the right area ($V_s$ = 2.0 V, I = 0.3 nA). Inset: line profile along the marked line in the STM image. (c) Atomically resolved STM image of 1T-PtSe$_2$ surface ($V_s$ = 0.3 V, I = 0.6 nA).

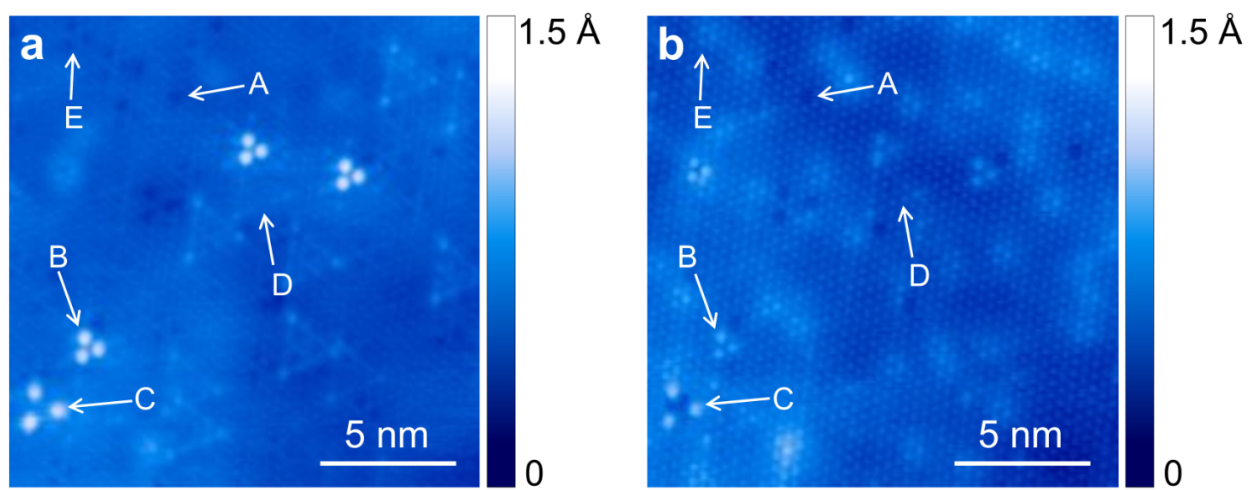

**FIG. 2.** STM micrographs of PtSe$_2$ film at 77K. (a) Empty-state image ($V_s$ = 0.4 V, I = 0.7 nA) and (b) Filled-state image ($V_s$ = − 0.4 V, I = 0.7 nA).

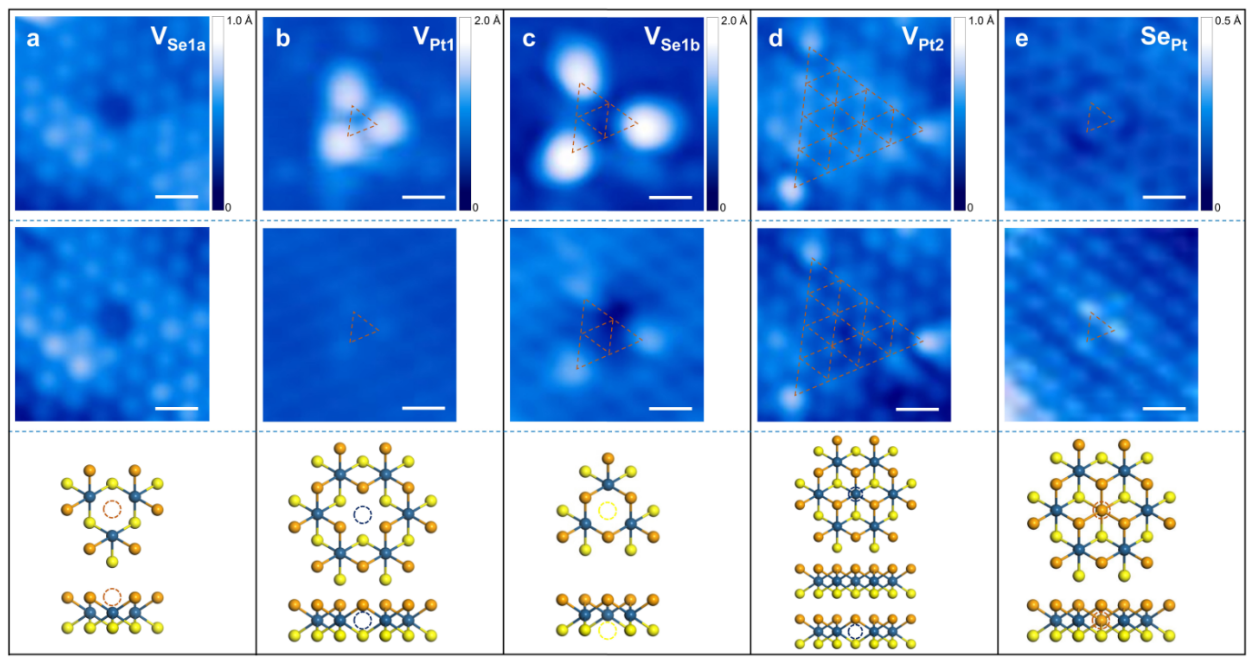

**FIG. 3.** Atomically resolved STM images of five types of defects. Top panel: Empty-state images. (a) Type A ($V_s = 0.3$ V, $I = 0.7$ nA). (b) Type B ($V_s = 0.2$ V, $I = 0.7$ nA). (c) Type C ($V_s = 0.2$ V, $I = 0.7$ nA). (d) Type D ($V_s = 0.085$ V, $I = 0.7$ nA). (e) Type E ($V_s = 0.3$ V, $I = 0.7$ nA). Middle panel: Filled-state images. (a) Type A ($V_s = -0.3$ V, $I = 0.7$ nA). (b) Type B ($V_s = -0.1$ V, $I = 0.7$ nA). (c) Type C ($V_s = -0.2$ V, $I = 0.7$ nA). (d) Type D ($V_s = -0.1$ V, $I = 0.7$ nA). (e) Type E ($V_s = -0.3$ V, $I = 0.7$ nA). The dashed lines indicate the size of defects in term of the lattice constant, appearing as 1×1, 2×2 and 3×3 triangles in (b), (c) and (d), respectively. Bottom panel: Top and side view of the models of each type of defects shown in the top and middle panels. The scale bar on the images is 0.5 nm.

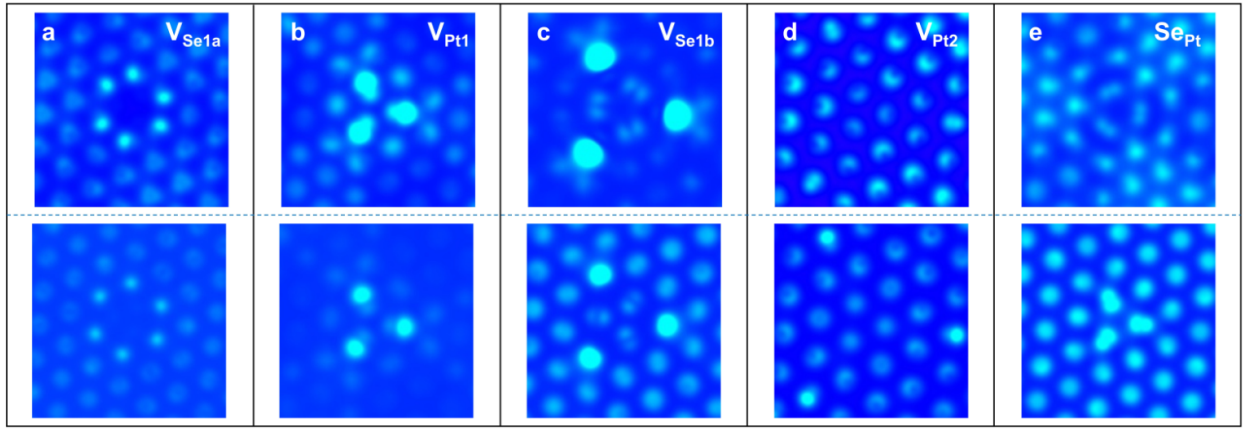

**FIG. 4.** DFT-calculated integrated local density of states for the five types of defects. Top panel: Empty-state images. (a) Type A at 0.25 eV. (b) Type B at 0.15 eV. (c) Type C at 0.2 eV. (d) Type D at 0.085 eV. (e) Type E at 0.15 eV. Bottom panel: Filled-state images. (a) Type A at − 0.3 eV. (b) Type B at − 0.1 eV. (c) Type C at − 0.2 eV. (d) Type D at − 0.1 eV. (e) Type E at − 0.3 eV.

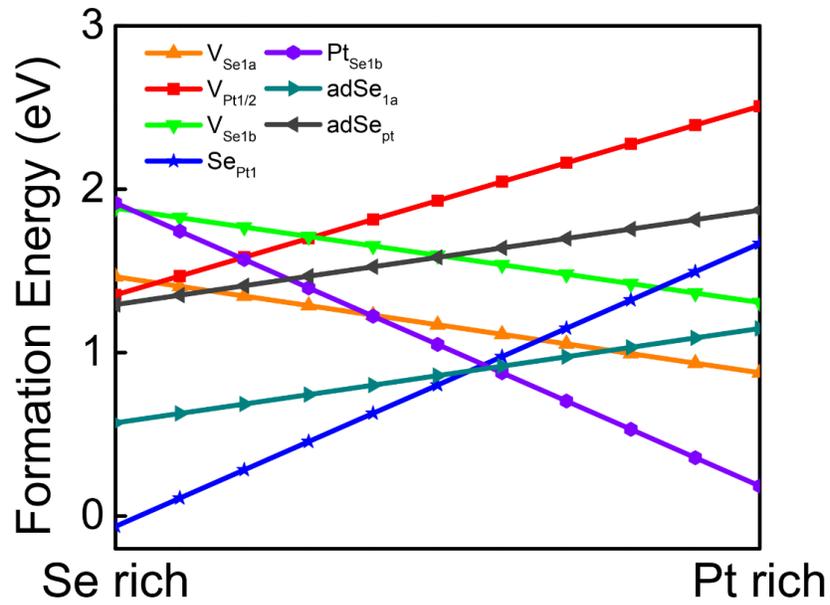

**FIG. 5.** Calculated defect formation energies. adSe$_{pt1}$ and adSe$_{1a}$ indicate Se adatoms at the hollow site above Pt atoms and the top site above Se$_{1a}$ atoms, respectively. The case of Se adatom at the hollow site above Se$_{1b}$ atoms gives a very similar result to the adSe$_{1a}$ case and not shown here. V$_{Pt1}$ and V$_{Pt2}$ have the same formation energies.

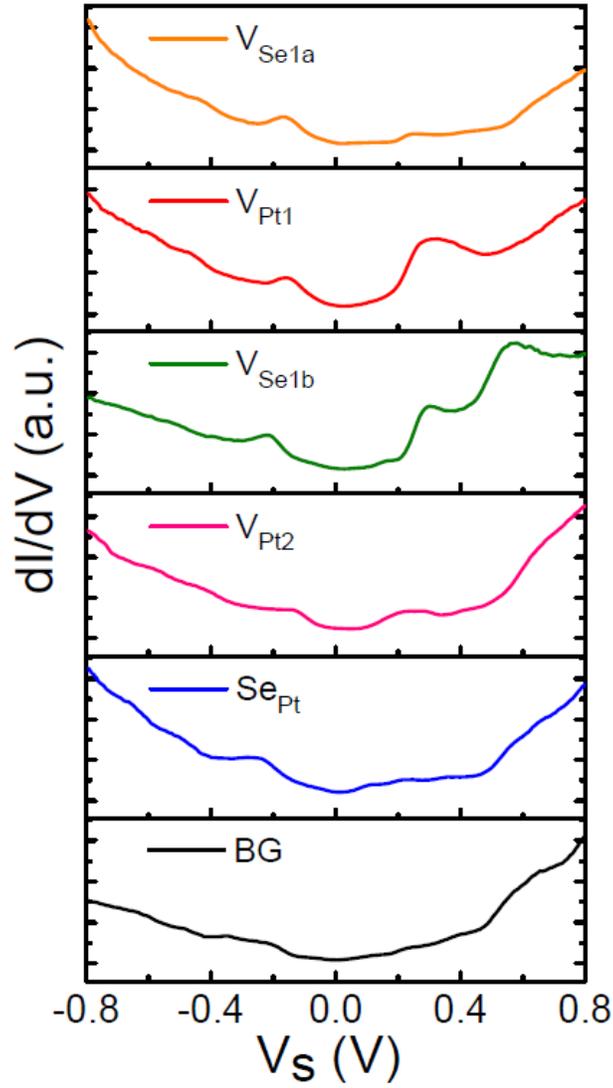

**FIG. 6.** STS of six-layer PtSe$_2$ with and without defects. The top five panels are for the layer with defects, while the bottommost panel is for the pristine layer.

**TABLE I.** Densities of defects obtained from STM and STEM measurements.

| Defects | Density obtained from STM images ($1/cm^2 \times 10^{12}$) |
| --- | --- |
| A ($V_{Se1}$) | $2.2 \pm 2.0$ |
| B ($V_{Pt1}$) | $1.2 \pm 0.4$ |
| C ($V_{Se2}$) | $1.2 \pm 0.6$ |
| D ($V_{Pt2}$) | $3.1 \pm 1$ |
| E ($Se_{Pt}$) | $16.4 \pm 3.9$ |